\documentclass[aps,floatfix,twocolumn,prl,superscriptaddress]{revtex4-2}
\usepackage{xcolor} 
\usepackage{graphicx}
\usepackage[hidelinks]{hyperref}
\usepackage{amssymb}
\usepackage[utf8]{inputenc}
\usepackage{amsmath}

\newcommand{\op}[1]{\hat{\mathrm{#1}}}

\renewcommand{\vec}[1]{\boldsymbol{#1}}
\newcommand{\iu}[0]{\mathrm{i}}

\begin{document}
\title{Electron magnetic moment of transient chiral phonons in KTaO$_3$}
\author{R. Matthias Geilhufe}
\affiliation{Department of Physics, Chalmers University of
Technology, SE-412 96 Gothenburg, Sweden}
\author{Wolfram Hergert}
\affiliation{Institute of Physics, Martin Luther University Halle-Wittenberg. D-06120 Halle, Germany}
\begin{abstract}
High intensity THz lasers allow for the coherent excitation of individual phonon modes. The ultrafast control of emergent magnetism by means of phonons opens up new tuning mechanisms for functional materials. While theoretically predicted phonon magnetic moments are tiny, recent experiments hint towards a significant magnetization in various materials. To explain these phenomena, we derive a coupling mechanism between the phonon angular momentum and the electron spin. This coupling introduces the transient level-splitting of spin-up and spin-down channels and a resulting magnetization. We estimate this magnetization on the example of the lowest infrared active mode in the perovskite KTaO$_3$. Our results show an electronic magnetic moment of $\approx 10^{-1}$~$\mu_B$ per unit cell, depending on the doping level and electron temperature. 
\end{abstract}
\maketitle
Chiral phonons or circularly polarized phonons carry angular momentum \cite{Zhang2014,mclellan1988angular}. As the collective excitation of charged ions, a magnetization occurs. Such a magnetization can be understood in the framework of the dynamical multiferroicity \cite{juraschek2017dynamical}, or, more specifically, the phonon inverse Farraday effect \cite{Juraschek2022,Juraschek2021,juraschek2020phono,rebane1983faraday}. However,
for a cyclotron motion, comparing the phonon magnetic moment $\mu_{\text{phonon}} = \frac{e \hbar}{2 m_{\text{ion}}}$ with the elecron magnetic moment $\mu_{B} = \frac{e \hbar}{2 m_{\text{e}}}$, the effect is smaller by a factor of $\frac{m_{\text{ion}}}{m_{\text{e}}}$, i.e., in the range of the nuclear magneton \cite{geilhufe2021dynamically,juraschek2019orbital}.

In contrast, recent experiments observed large phonon magnetic moments. In such experiments, optical phonons are excited with a circularly polarized THz laser pulse. Cheng \textit{et al.} measured the phonon Zeeman effect in the Dirac semimental Cd$_3$As$_2$ and observed a moment of $\approx 2.7~\mu_B$ \cite{cheng2020large}. They explain the giant magnetic moment by a coupling of the phonon to the effective Dirac electron, in resonance with the cyclotron frequency. Baydin \textit{et al.} revealed a phonon Zeeman splitting of similar size in the semiconductor PbTe \cite{baydin2022magnetic}. Here, the magnetic moment was explained in terms of strong anharmonicity effects. The effect even increases after a topological phase transtion in thin films of the related compund Pb$_{1-x}$Sn$_x$Te ($x>0.32$) \cite{hernandez2022chiral}. Basini \textit{et al.} measured the dynamical magnetization by chiral phonons in SrTiO$_3$ using optical Kerr rotation, and, again found a magnetic moment of similar size \cite{basini2022}. The effect is explained by the angular momentum transfer from phonons to electrons in terms of the inverse Barnett effect and a resulting enhanced gyromagnetic ratio. 

The mismatch of theory and experiment gives evidence that phonons cannot be described in the absence of electrons. Instead, electron-phonon and, specifically, spin-phonon coupling play a significant role in explaining the emergent dynamical magnetization. Mechanisms for interactions of the lattice and spins have been known for a long time, e.g., from the early work of Elliot and Yafet on spin relaxation \cite{elliot1954,yafet1963solid} and have let to sophisticated ab initio implementations \cite{Fransson2017}. Generally, the coupling of lattice degrees of freedom to the spin requires spin-orbit interaction, and can be modeled by assuming small atomic displacements in the potential energy \cite{Fransson2020}. Other approaches are based on modifications in the magnetic exchange \cite{son2019unconventional,weber2022emerging}, or an inverse Katsura-Nagaosa-Balatsky mechanism \cite{KNB2005,Mostovoy2006,Mochizuki2011}. To incorporate an additional orbital magnetization, Ren \textit{et al.} \cite{Ren2021} showed that the second Chern form times the phonon angular momentum gives rise to a magnetization, in the framework of the modern theory of magnetization \cite{Xiao2005,Thonhauser2005,resta2010electrical}. Additionally, a non-zero phonon angular momentum might also induce dynamical coupling terms based on quantum inertial effects \cite{geilhufe2022dynamic}, similar to the expected spin-rotation coupling in mechanical resonators inducing accoustinc phonons \cite{Matsuo2013,Matsuo2011,Matsuo2011b}. The expected size of a dynamical magnetization due to spin-rotation coupling will be discussed throughout this paper. 

For simplicity, we consider a two-fold degenerate phonon mode $\vec{q}=(q_1,q_2)$ (in units \AA $\sqrt{u}$, with $u$ the atomic mass unit) with eigenfrequency $\omega_0$. The phonon Lagrangian is given by \cite{juraschek2017dynamical}
\begin{equation}
    \mathcal{L}^{\text{phonon}}(\vec{q},\dot{\vec{q}}) = \frac{\dot{\vec{q}}^2}{2} - \frac{\omega_0^2\vec{q}^2}{2}.
    \label{eq:L-phonon}
\end{equation}
The electron Lagrangian of a non-relativistic electron in a nominally non-magnetic material and in the absence of external electromagnetic fields is given by
\begin{multline}
    \mathcal{L}^{\text{electron}}(\vec{\Psi},\vec{\Psi}^\dagger)\\ = \int \mathrm{d}^3r \left[\iu\hbar \vec{\Psi}^\dagger \dot{\vec{\Psi}} -  \vec{\Psi}^\dagger\left(\frac{\op{\vec{p}}^2}{2m} + V(\vec{r})\right)\vec{\Psi}\right].
    \label{eq:L-electron}
\end{multline}
We formulate the spin-phonon coupling term with symmetry arguments, by considering the two fundamental symmetries time-reversal $\mathcal{T}$ and spatial inversion or parity $\mathcal{P}$. The phonon mode $\vec{q}$ describes the displacement of ions from their equilibrium position. As a result, it transforms as a vector, being odd under $\mathcal{P}$ and even under $\mathcal{T}$. It follows that the time-derivative $\dot{\vec{q}}$ is odd under both $\mathcal{P}$ and $\mathcal{T}$. This is consistent with the transformation behavior of the phonon angular momentum $\vec{L}^{\text{phonon}} = \vec{q}\times\dot{\vec{q}}$, being a pseudo vector, i.e., even under $\mathcal{P}$ and odd under $\mathcal{T}$. The same holds for the electron total angular momentum $\vec{J}^{\text{electron}} = \vec{L}^{\text{electron}} + \vec{S}^{\text{electron}}$, with $\vec{L}^{\text{electron}}$ being the orbital angular momentum and $\vec{S}^{\text{electron}}$ the spin. Hence, we can write down a $\mathcal{P}$ and $\mathcal{T}$ invariant scalar 
\begin{equation}
  \mathcal{L}^{\text{coupl.}} = -\alpha \vec{L}^{\text{phonon}}\cdot \left<\vec{J}^{\text{electron}}\right>,
   \label{eq:L-epcoupling}
\end{equation}
with $\alpha$ being a coupling constant in units of the inverse of the moment of inertia. Note, that we use a semiclassical approach here, with the phonon dynamics described classically. This is because we assume a non-equilibrium state with strong phonon amplitude. Hence, the phonon angular momentum is an actual angular momentum \cite{tauchert2022polarized}, as described towards the end of Ref. \cite{streib2021difference}. By expressing the phonon angular momentum in terms of the angular velocity $\vec{\omega}^{\text{phonon}}$ and the moment of inertia $I$, equation \eqref{eq:L-epcoupling} can be brought into the form of the spin-rotation coupling discussed in Refs. \cite{geilhufe2022dynamic,Matsuo2013,Matsuo2011,Matsuo2011b}, $\mathcal{L}^{\text{coupl.}} = \vec{\omega}^{\text{phonon}}\cdot \left<\vec{J}^{\text{electron}}\right> = -I^{-1} \vec{L}^{\text{phonon}}\cdot \left<\vec{J}^{\text{electron}}\right>$. The moment of inertia is given by $I = \int \mathrm{d}\vec{r} \vec{r}^2 m(\vec{r})$, with $m(\vec{r})$ the mass density. 

\begin{figure}
    \centering
    \includegraphics[width=\linewidth]{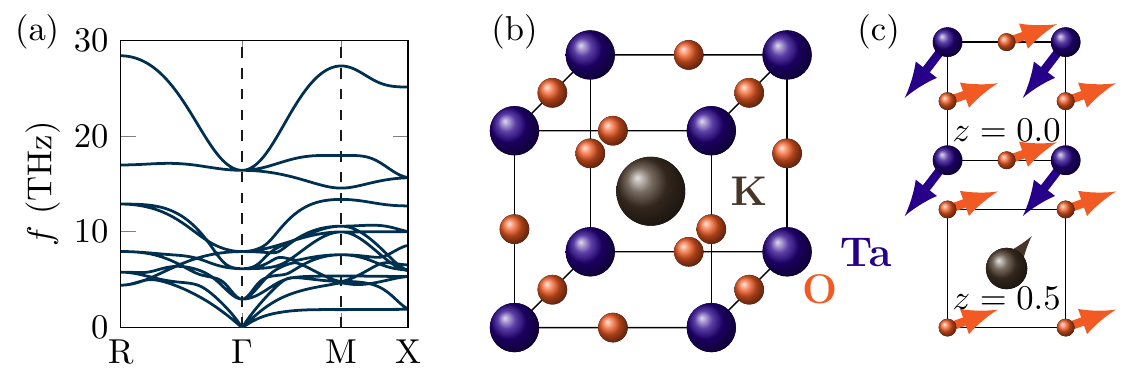}
    \caption{Phonons in KTaO$_3$. (a) Full phonon spectrum; (b) crystal structure; (c) eigenvector of the lowest infrared active mode.}
    \label{phonons}
\end{figure}

We continue by evaluating the spin-rotation couplig on the example of KTaO$_3$. 
The phonon dynamics is described using classical equations of motion and coupling the two-fold degenerate phonon normal mode $\vec{q}$ to a circularly polarized THz laser pulse $\tilde{\vec{E}}(t)$,
\begin{equation}
  \ddot{\vec{q}}(t) +\eta \dot{\vec{q}}(t) + \omega^2_0 q_i(t) = \tilde{Z} \tilde{\vec{E}}_0(t) .
  \label{eqm}
\end{equation}
Here, $\eta$ is the damping or line width and $\omega_0$ the mode frequency. We focus on the temperature dependent soft mode, which, at $300~\text{K}$ has $\omega_0= 2.42$ and $\eta = 0.64$ \cite{Voigt1995SrTiO3KTaO3}. $\tilde{Z}$ is the mode effective charge. For KTaO$_3$ it takes the value $\tilde{Z} = 1.4$ \cite{juraschek2019orbital}. For the pulse we assume $\tilde{\vec{E}} = \tilde{E}_0 \exp\left(\frac{1}{2}\frac{(t-t_0)^2}{\tau}\right)(\sin(2\pi\omega t),\cos(2\pi\omega t))$. $\tilde{E}_0 = \epsilon_\infty^{-1}E_0$ is the screened field strength with $\epsilon_\infty = 4.3$ \cite{Barker1964}, where we choose $E_0 = 1~\text{MVcm}^{-1}$.  The pulse width is chosen to be $\tau = \sqrt{\frac{1}{2}}~\text{ps}$. For simplicity, we neglect higher-order couplings \cite{kozina2019terahertz}. 

The full phonon spectrum of KTaO$_3$ as well as one eigenvector of the soft mode is shown in Fig. \ref{phonons}. The other two Eigenmodes are obtained by applying cubic symmetries. The eigenvectors and phonon spectrum have been obtained using density functional perturbation theory \cite{king1993theory,gajdovs2006linear} as implemented in VASP \cite{vasp} and Phonopy \cite{togo2015first}. The exchange correlation functional was approximated by the PBE generalized gradient approximation \cite{perdew1996generalized}. We used a $\vec{k}$-mesh density of $\approx 1050\,\vec{k}\text{-points}/\text{\AA}^{-3}$, i.e., an $8\times 8\times 8$ mesh and a cut-off energy of $700$ eV.

\begin{figure}
    \centering
    \includegraphics[width=\linewidth]{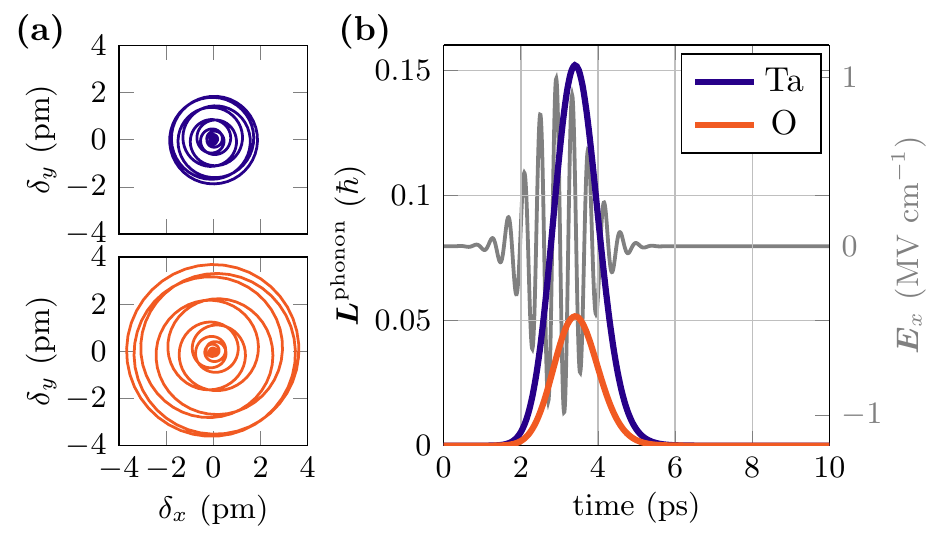}
    \caption{Calculated atomic motion of oxygen and tantalum in the damped circularly polarized phonon excitation. (a) Displacement $\vec{\delta}$; (b) site-resolved phonon angular momentum and laser pulse in resonance with the lowest infrared active mode of KTaO$_3$. The damping is $\eta = 0.1\times 2\pi~$THz. }
    \label{phononnumerics}
\end{figure}

The numerical solution of equation \eqref{eqm} is shown in Fig. \ref{phononnumerics}. We focus on O and Ta, as they are the relevant atoms in the electronic structure discussed below. The largest displacement from the atomic equilibrium position is shortly after the pulse peak at 3~ps. The large damping of the phonon mkes a distinction from the inverse Faraday effect, i.e., the DC magnetization of a sample exposed to circularly polarized light, challenging. The inverse Faraday effect originates from the time-reversal symmetry breaking due to the photon angular momentum, a term which couples to the magnetic field in the free energy \cite{pershan1966theoretical}. In contrast, the phonon inverse Faraday effect emerges due to circularly polarized phonons and does not require the presence of an external electric field. Instead, a magnetization is also predicted for thermally excited chiral phonons \cite{Hamada2020,gurtubay2020,li2021chiral}.

\begin{figure}
    \centering
    \includegraphics[width=\linewidth]{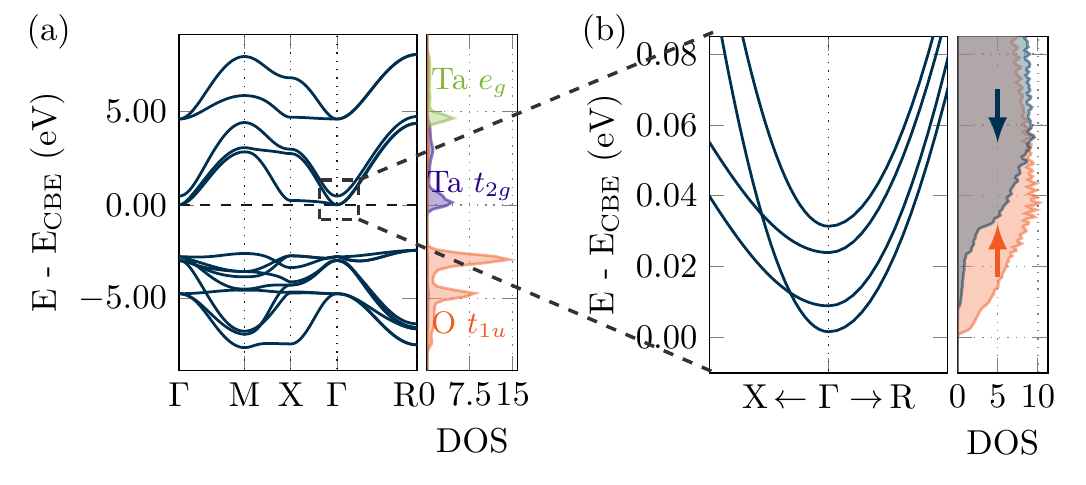}
    \caption{Electronic structure of KTaO$_3$ relative to the conduction band edge ($E_{\text{CBE}}$). (a) Full band structure. (b) Expected band splitting of the conduction band due to spin-rotation coupling close to the $\Gamma$ point. The DOS is projected on spin-up and spin-down channels. }
    \label{bands}
\end{figure}

The electronic structure of KTaO$_3$ is modeled in a two-center Slater-Koster tight-binding scheme \cite{podolskiy2004compact,slater1954simplified} as implemented in the Mathematica group theory package GTPack \cite{gtpack1,gtpack2} (cf. Appendix). The band structure close to the Fermi energy is plotted in Fig. \ref{bands}(a). The valence bands are mainly composed of the O-$p$ orbitals, whereas the conduction band is dominated by contributions from Ta-$d$ orbitals. The splitting into $t_{2g}$ and $e_g$ bands due to cubic symmetry is clearly revealed. The band gap is $\approx 2.8$~eV~ and lower than the experimentally observed optical band gap of $\approx 3.6$~eV \cite{Jellison2006}. The splitting of the $t_{2g}$ band at $\Gamma$ due to spin-orbit interaction is $\approx 0.46$~eV.

\begin{figure}
    \centering
    \includegraphics[width=4.5cm]{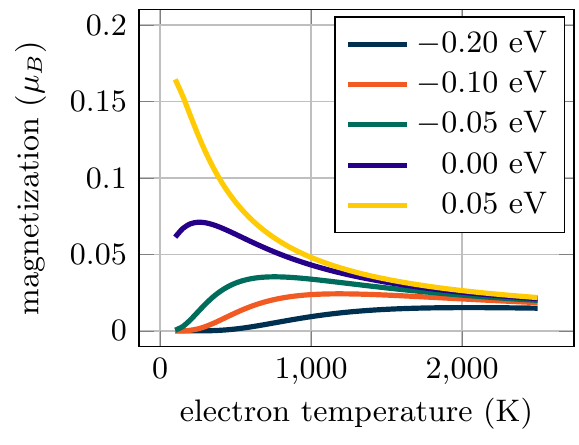}
    \caption{Electronic magnetization due to spin-phonon interaction. Estimates are given for various values of the chemical potential, relative to the conduction band minimum. }
    \label{magnetization}
\end{figure}

In the presence of transient chiral phonons, the time-reversal symmetry of the system is broken. As a result, Kramers degeneracy of the electronic bands is lifted. To model this scenario, the tight-binding Hamiltonian is extended by the spin-phonon coupling of equation \eqref{eq:L-epcoupling}. We choose $I^{-1}\vec{L}^{\text{phonon}}=\vec{\omega}$ \cite{geilhufe2022dynamic}, with $\vec{\omega}$ being the angular velocity of the site-resolved displacement having the frequency $\omega_0 = 2.42\times 2\pi$~THz. The corresponding coupling to the total angular momentum of the electron, $\vec{\omega}\cdot\vec{J}$ is diagonal in the $\left|j,\mu\right>$ basis. We project the resulting band structure onto spin-up and spin-down channels as shown in Fig. \ref{bands}(b). We calculate the magnetization by
\begin{equation}
    M_z = \mu_B \left(N_\uparrow-N_\downarrow\right),~ N_\sigma = \int_{-\infty}^{\infty} \mathrm{d}E\,n_\sigma(E) f(E,\mu,T),
    \label{Mageq}
\end{equation}
with $n_\sigma(E)$ the electronic density of states of spin $\sigma$ and $f(E,\mu,T)$ the Fermi-Dirac distribution. 
The magnetization \eqref{Mageq} is computed with the tight-binding band structure in Fig. \ref{bands}(b). The dynamically induced magnetization is shown in dependence of the chemical potential $\mu$ and electron temperature $T$ in Fig. \ref{magnetization}. As can be seen, an electronic magnetization of $\approx 10^{-1}~\mu_B$ can be achieved. This moment is about three orders of magnitude larger than the estimate for a purely ionic effect in KTaO$_3$ \cite{geilhufe2021dynamically}.

In the estimates, we focus on the chemical potential close to the conduction band edge. In fact, intrinsic doping in KTaO$_3$ crystals, e.g., due to O-vacancies induce sub-gap states originating from $d$-electrons of Ta atoms close to the O vacancy \cite{Choi2011,modak2021energetic}. While a natural amount of doping ($< 10^{18}$ charge carriers/cm$^{3}$) still leads to an insulating behavior, a larger amount of O vacancies can induce a metallic state \cite{ueno2011discovery}. In the insulating state, a fairly large magnetization occurs if the electronic temperature becomes high. Such a situation might occur for short durations after the crystal was exposed to the laser light \cite{alber2021ntmpy,unikandanunni2022ultrafast}. Due to the reduced mass of electrons compared to ions, the kinetic energy will be significantly larger until it relaxes into an equilibrium state together with the ions. In contrast, if the Fermi level slightly cuts the conduction band, a lower electronic temperature would be required to exclude contributions to the magnetization from the minority spin-channel. Also, we note that electrons can be excited from the sub-gap states into the conduction band by a multiphoton process in intense laser fields. For low photon energy THz fields, the photon density is significantly higher, making a multi-photon process more likely. Also, such a mechanism can lead to the formation of excitons and induce the transient shift of the chemical potential for electrons and holes. 

In summary, we estimated the effect of the chiral spin-phonon coupling $\sim \vec{L}^{\text{phonon}}\cdot\vec{J}^{\text{electron}}$. This mechanism induces a split of electronic spin states in the adiabatic approximation. As a consequence, a mismatch of occupied spin-up and spin-down states leads to an emergent magnetization in the order of $10^{-1}~\mu_B$. Hence, the effect is larger than the purely ionic phonon inverse Faraday effect by three orders of magnitude. The chiral spin-phonon coupling might be relevant to explain the large experimentally observed phonon magnetic moments due to chiral phonons in various materials \cite{cheng2020large,baydin2022magnetic,basini2022,hernandez2022chiral}. Furthermore, also the opposite effect, i.e., the transfer of electronic angular momentum to the lattice can be described by the chiral spin-phonon coupling and would contribute, e.g., to ultrafast demagnetization effects \cite{tauchert2022polarized,Tsatsoulis2016,koopmans2010explaining}. 

{\it Acknowledgment.} We are grateful for inspiring discussions with A. V. Balatsky, S. Bonetti, M. Basini, D. Juraschek, G. Fiete, M. Rodriguez-Vega, and A. Ernst. RMG acknowledges support from Chalmers University of Technology. Computational resources were provided by the Swedish National Infrastructure for Computing (SNIC) via the High Performance Computing Centre North (HPC2N) and the Uppsala Multidisciplinary Centre for Advanced Computational Science (UPPMAX).

\section*{Appendix}
We construct the tight-binding Hamiltonian using the Slater-Koster scheme \cite{podolskiy2004compact,slater1954simplified} as implemented in the Mathematica group theory package GTPack \cite{gtpack1,gtpack2}. A restricted basis set containing O $p$-orbitals and Ta $d$-orditals is used. Spin-orbit coupling of Ta $d$-orbitals is included. The Hamiltonian in two-center approximation considers nearest neighbor interactions of Ta and nearest and next-nearest neighbor interactions of O. The Hamiltonian is represented by a $28\times 28$ matrix and contains 14 parameters (s. Tab. \ref{parm}). The crystal field splitting for the on-site parameters is taken into account.
\begin{table}[h]
\caption{Tight-binding parameters of the Hamiltonian \label{parm}}
 \renewcommand{\arraystretch}{1.5}
 \tiny
\begin{tabular}{|c|c|c|c|c|c|c|}
\hline
$(dd\sigma)_1^\textrm{Ta,Ta}$ &$(dd\pi)_1^\textrm{Ta,Ta}$ &$(dd\delta)_1^\textrm{Ta,Ta}$ &$(pp\sigma)_1^\textrm{O,O}$&$(pp\pi)_1^\textrm{O,O}$&$(pp\sigma)_2^\textrm{O,O}$&$(pp\pi)_2^\textrm{O,O}$\\
\hline
0.11409&-0.24820 &0.02087&0.38702&-0.0955& 0.14458 &-0.042634 \\
\hline
$(pd\sigma)_1^\textrm{0,Ta}$ &$(pd\pi)_1^\textrm{O,Ta}$&$(dd0)_\textrm{t2g}^\textrm{Ta} $ &$(dd0)_\textrm{eg}^\textrm{Ta} $&  $(pp0)_\perp^\textrm{O} $ &$(pp0)_\parallel^\textrm{O}$ &  $\xi$\\
\hline
-2.77242& 1.57614 & 3.08778 & 6.15932 &-1.4078 &-2.44059 & 0.28868\\
\hline
\end{tabular}
\end{table}

\begin{figure}[h!]
    \centering
    \includegraphics[width=4.5cm]{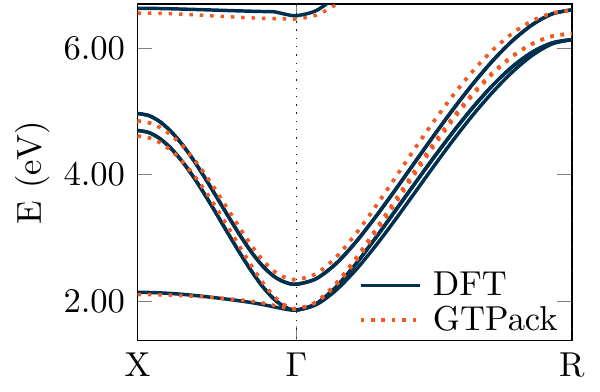}
    \caption{Comparison of the DFT band structure and the 2-center tight-binding band structure using GTPack. }
    \label{Appendix:DFTfit}
\end{figure}

The tight-binding Hamiltonian is fitted to DFT band structure data (computational details given in the main text). We refine an initial guess by the least squares method. 15 $\vec{k}$-points are taken along each symmetry line. To avoid the algorithm getting stuck in a local minimum in parameter space, we add noise to the parameters after a first step and repeat the minimization scheme. The final result is shown in Fig. \ref{Appendix:DFTfit}. It can be seen that the relevant conduction band is well-parametrized close to the $\Gamma$ point.

%
\end{document}